# Single-Crystal N-polar GaN *p-n* Diodes by Plasma-Assisted Molecular Beam Epitaxy

YongJin Cho,[1,a] Zongyang Hu,[1] Kazuki Nomoto,[1] Huili Grace Xing,[1,2] and Debdeep Jena[1,2,b]

[1]*School of Electrical and Computer Engineering, Cornell University, Ithaca, New York 14853, USA*
[2]*Department of Materials Science and Engineering, Cornell University, Ithaca, New York 14853, USA*

ABSTRACT

N-polar GaN *p-n* diodes are realized on single-crystal N-polar GaN bulk wafers by plasma-assisted molecular beam epitaxy growth. The current-voltage characteristics show high-quality rectification and electroluminescence characteristics with a high on currents ~10 kA/cm$^2$, low off currents <10$^{-5}$ A/cm$^2$, on/off current ratio of >10$^9$, and interband photon emission. The measured electroluminescence spectrum is dominated by strong near-band edge emission, while deep level luminescence is greatly suppressed. A very low dislocation density leads to a high reverse breakdown electric field of ~2.2 MV/cm without fields plates – the highest reported for N-polar epitaxial structures. The low leakage current N-polar diodes open up several potential applications in polarization-engineered photonic and electronic devices.

---

[a] Electronic mail: yjcho@alumni.nd.edu
[b] Electronic mail: djena@cornell.edu

Wurtzite III-Nitride semiconductor heterostructures exhibit strong spontaneous and piezoelectric polarization fields of the order of a few MV/cm along the polar *c*-axis.[1] These polarization fields cause quantum-confined Stark effect in the active regions of quantum-well light emitting devices. The polarization-induced reduction of the oscillator strength due to poor electron–hole overlap is thought to reduce the efficiency of LEDs, generating interest in growth along non-polar and semi-polar directions.[2,3] On the other hand, the built-in polarization fields can be advantageous in generating 2D electron gases in high-electron mobility transistors (HEMTs), and for polarization induced *p*-type doping.[4,5] Polarization engineering in heterostructures offers several new opportunities for photonic and electronic devices, tunnel junctions, including ultra-low power tunneling transistors.[6,7]

Most studies of nitride electronic devices so far have focused on heterostructures grown in the metal-polar (Ga- or Al-polar) directions. The opposite direction of polarization, i.e., N-polar direction, can also be employed in growth for unique device properties such as buried-barrier HEMTs and interband tunnel junctions.[8,9,10,11] Epitaxial growth along the N-polar direction presents certain fundamental advantages stemming from the polarity-dependent decomposition temperatures of the materials. This advantage enables N-polar growth at much higher temperatures than the cation-polar counterparts.[11,12,13] Yet, homoepitaxy along the N-face orientation has received little attention due to lack of bulk substrates, coupled with difficulties of high-quality epitaxy on defective substrates with high density of dislocations and rough surfaces.

In this work, by taking advantage of newly available N-polar single crystal GaN bulk substrates with high structural perfection and atomically flat surfaces, we demonstrate that high-quality N-polar GaN *p-n* diodes comparable to the state-of-the-art Ga-polar counterparts can be obtained by plasma-assisted molecular beam epitaxy (PA-MBE). The resulting homoepitaxial *p-n* junction diodes exhibit off-state leakage current <$10^{-5}$ A/cm$^2$ and a peak electric field of 2.2 MV/cm at a breakdown voltage, making them the highest quality *p-n* diodes ever demonstrated on N-polar GaN.

Epi-ready N-polar n$^+$-type (000-1) GaN with free electron concentration ~$10^{19}$ cm$^{-3}$ produced by Ammono SA were used as the starting substrates. Atomic force microscopy (AFM) measurement revealed a typical root-mean-square (rms) roughness of the N-polar epi-ready surface of ~0.4 nm over 10 × 10 μm$^2$. X-ray diffraction rocking curve across the (0002) reflection exhibited a full-width-at-half-maximum of 20 arcsec and a dislocation density of ~5 × 10$^4$ cm$^{-2}$, orders of magnitude lower than the conventional ~$10^9$/cm$^2$ values on alternative substrates. The GaN *p-n* diodes were directly grown on the N-polar GaN(000-1) surface in a Veeco Gen Xplor MBE chamber equipped with standard effusion cells for Ga, Si and Mg, and a radio frequency plasma source for active N species. Si was used as the n-type donor, and Mg as the *p*-type acceptor to realize the conductive layers of the GaN *p-n* diode. The base pressure of the growth chamber was <$10^{-10}$ Torr under idle conditions, and ~7×$10^{-5}$ Torr during the growth runs. As shown in Fig. 1, the MBE-grown *p-n* diodes layer structure starting from the nucleation surface is: 100 nm GaN:Si / 400 nm GaN / 150 nm GaN:Mg / 20 nm GaN:Mg. The unintentionally doped (uid) 400 nm-thick GaN layer shows *n*-type conductivity due to unintentional O incorporation.[14] The heavily doped top GaN:Mg layer was grown to facilitate low-resistance *p*-type Ohmic contacts. The doping concentrations of the layers are shown in Fig. 1(c). All these layers were grown under Ga–rich condition at a growth rate of 840 nm/h. The Si doped- and uid-GaN layers were grown at a substrate temperature of 700 °C, which was lowered to 630 °C for the *p*-GaN, and 580 °C for the *p*$^+$-GaN layers to obtain the intended Mg doping concentrations.[15] Both Si and Mg doping concentrations were calibrated in advance with secondary-ion mass spectroscopy measurements on separate doping calibration growth stacks (not shown here). The excess Ga droplets after the growth were first removed in HCl before *ex situ* characterizations and device fabrication.

The morphology and structural quality of the samples were evaluated by *in situ* reflection high energy electron diffraction (RHEED), AFM, and transmission electron microscopy (TEM). Figure 1(a) shows a RHEED pattern along the <11-20> azimuth of the N-polar GaN *p-n* diode surface measured after cooldown to <200 °C after growth. It reveals pure reflection patterns with a well-defined specular spot and pronounced Kikuchi lines, indicating smooth surface morphology and high structural order. More importantly, a clear (3×3) surface reconstruction, the fingerprint of a Ga adlayer on an N-polar GaN surface, is clearly observed, confirming the N-polarity of GaN *p-n* diodes.[16] For the PA-MBE growth of GaN, a metallic Ga adlayer is beneficial to enhance the migration length of adatoms.[12,17] Here, it enabled extremely smooth surfaces after the MBE growth, exhibiting clear atomic steps with rms roughness of ~0.35 nm as shown in Fig. 1(b).

In order to fabricate the *p-n* junction devices, the epitaxially grown samples were first cleaned using solvents and HF. For the *n*-type contact, 50/100 nm Ti/Au stacks were deposited on the backside of the substrate. Then 50/100/50 nm Pd/Au/Ni were deposited on the top GaN:Mg layer for *p*-type Ohmic contacts through a lithographic mask. Before mesa etching, circular transmission line model (C-TLM) analysis on the *p*-layers were performed to extract the *p*-type contact resistivity (~$5.1\times10^{-3}$ $\Omega cm^2$) and sheet resistance (~$1.4\times10^5$ $\Omega$/sq). Using the *p*-contacts as etch masks for device isolation, >600 nm tall mesas were formed by reactive ion etching. A schematic of the processed device is displayed in Fig. 1(c). Figure 1(d) shows a cross-sectional TEM image of a fully processed N-polar GaN *p-n* diode. There is a sharp distinction between the top *p*-contact metal stack and the epitaxial GaN crystal: contrary to the defects and inhomogeneities in the *p*-contact metal layers, no extended defects such as threading dislocations is observed from the whole MBE-grown GaN region in the micrograph. The entire GaN epilayer and substrate appear as a single bulk crystal with no interfacial features either. 2-beam bright-field cross-sectional TEM on the devices using diffraction vectors (0001) and (1-100) revealed no dislocations within the probed areas (not shown here). These observations indicate that the high structural perfection of the single-crystal GaN substrate was successfully largely transferred to the MBE-overgrown *p-n* diodes.

We now turn to the electronic and photonic properties of the N-polar GaN *p-n* diodes as seen from current-voltage (*I-V*), junction capacitance-voltage (*C-V*), and electroluminescence (EL) measurements. These *ex situ* characterizations were all performed under ambient conditions at 300 K. Figure 2(a) depicts in logarithmic scale the *I-V* characteristics of diodes of diameter ~20 µm, exhibiting high quality rectification, a hallmark of high-quality GaN *p-n* diodes.[18,19] The inset in Fig. 2(b) shows the linear *I-V* characteristics of a 50 um diode, with a near band-edge *p-n* diode turn-on voltage of ~3.5 V. The forward current density reaches ~7.8 kA/cm$^2$ at 5 V, from which a differential on-resistance $R_{ON}$~0.1 mΩ·cm$^2$ is extracted, which includes the probe resistance (the real $R_{ON}$ is smaller). The leakage current density of the diodes remains lower than 10$^{-5}$ A/cm$^2$ limited by the experimental setup for bias voltages from -6 V to +1 V and the on/off current ratio at ±5 V is >10$^9$. This high performance proves high-quality N-polar GaN *p-n* diodes with a low dislocation density are now possible by MBE.

Carriers injected across *p* and *n* regions in a *p-n* diode recombine by two mechanisms: radiative and nonradiative. Dislocations are predominant nonradiative centers for carrier recombination and point defects such as vacancies are deep levels act as deep luminescence centers in GaN.[20] In order to evaluate the radiative properties of the diodes, especially to assess the population of the point defect-related deep levels, EL measurements were performed under forward bias. Figure 2(b) shows the measured EL spectrum for an N-polar GaN *p-n* diode with a diameter of 50 μm at 5 V and a current injection of ~1.5 kA/cm$^2$. The current-voltage relation of this device is shown in the inset of Fig. 2(b). Two EL peaks at 3.13 eV and 3.39 eV are seen. The peak at 3.39 eV is very close to the fundamental bandgap of GaN and is assigned to the interband $E_g = E_c - E_v$ or Near Band-Edge (NBE) transition. The energy separation between the two peaks is very close to the Mg acceptor binding energy ($E_a$ ~0.25 eV) in GaN,[21] and therefore the peak at 3.13 eV is believed to be due to $E_c - (E_v + E_a) = E_g - E_a$, the conduction band-to-acceptor transition (CBA) in the GaN:Mg layers. A weaker, broad luminescence band centered at ~2.2 eV is also seen. This band is due to deep-level transitions (DL) by point defects in GaN.[20,22] Note that the intensity of this broad luminescence band is much weaker than the other two peaks and the spectrum is dominated by the >3 eV near band edge emissions. The presence of the NBE and CBA peaks, and the weak intensity of the broad band peak indicate a low density of deep point defects in the *p-n* diodes.

We next study the breakdown behavior of the N-polar *p-n* diodes. Figure 3(a) shows the reverse-bias *I-V* characteristics of a 20 μm diameter diode. With increasing reverse bias voltage, the current density gradually increases from <10$^{-5}$ A/cm$^2$ (limited by the experimental setup) to 10$^{-1}$ A/cm$^2$ and abrupt electrical breakdown occurs at $V_{br}$~ - 76 V. The magnitude of the reverse-bias current density suggests that the reverse bias current before breakdown is likely due to the trap-assisted avalanche effects and *not* interband Zener tunneling.[23,24]

In order to estimate the doping concentration in the uid-GaN layer, *C-V* measurements were performed on 30 µm diameter diodes at a frequency of 1 MHz, as shown in Fig. 3(b). A loss tangent angle >82° indicates low leakage in the voltage range of measurement. The corresponding $1/C^2$ vs *V* plot and a linear fit shown in Fig. 3(b) allows us to extract an uid-region donor doping density $N_d=9.6\times10^{16}$ cm$^{-3}$ for the diodes, assuming acceptor doping on the *p*-side $N_a \gg N_d$. The voltage-intercept of the linear fit gives a built-in voltage of 3.3 V, close to the theoretical value of 3.2 V obtained with the materials parameters in Fig. 1(c). Using this uid-region doping density, we calculate the electric field profile at the breakdown voltage by solving the Poisson equation and show the result in Fig. 3(c). Since $N_a \gg N_d$, the depletion region is located mostly in the *n*-side of the diode, while the uid-GaN region is completely depleted, as can be seen in Fig. 3(c). At this breakdown voltage, a peak electric field of 2.2 MV/cm at the edge of the depletion region is estimated. This breakdown electric field, lower than the best Ga-polar GaN *p-n* diodes of ~4 MV/cm,[19] nevertheless indicates the highest value for N-polar GaN *p-n* diodes and can be significantly improved by sculpting the electric field externally using field-plates as the Ga-polar counterparts.[19,25,26] The full performance and true breakdown behavior of the diodes may be accessible by electrically isolating the device regions from edge sidewalls.[27]

In summary, N-polar GaN *p-n* vertical diodes of high quality were demonstrated on single-crystal N-polar GaN wafers by PA-MBE growth. The overgrown epitaxial N-polar *p-n* diodes were seen to follow the high structural perfection of the underlying high-quality N-polar GaN substrates. Single-crystal GaN substrates with very low dislocation densities enable low leakage currents and the high breakdown field in these diodes. N-polar epitaxial GaN *p-n* diodes are fundamentally distinct from a Ga-polar *p-n* diode that is simply turned upside down; because once InGaN or AlGaN heterostructures are incorporated in the depletion region or elsewhere, the polarization-induced fields point in opposite directions. For various photonic and electronic devices, it is desired that the *p*-layer be on the top, and that forces devices such *p*-side up tunneling transistors to be built on N-polar substrates. The work presented in this work shows that high-quality N-polar GaN *p-n* diodes are now feasible by MBE growth as the first step towards the realization of a number of new device possibilities.

The authors thank the Notre Dame Integrated Facility for the help with TEM. This work was supported by the Center for Low Energy Systems Technology (LEAST), one of the six SRC STARnet Centers, sponsored by MARCO and DARPA.

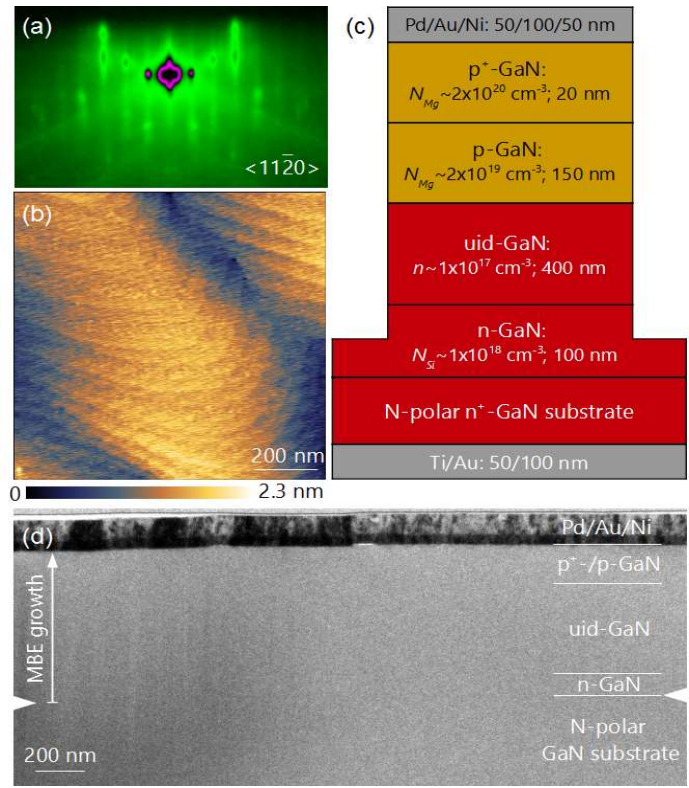

Fig. 1 (a) RHEED pattern showing the 3 × 3 reconstruction characteristic of N-polar surface, and (b) AFM micrograph of MBE-grown N-polar GaN *p-n* diodes showing atomic steps. The RHEED pattern was taken below 200℃ along the <11-20> azimuth after growth. (c) Schematic layer structure and (d) cross-section TEM micrograph of the fabricated vertical *p-n* diodes. The two white grooves on the sides of the TEM image highlight the interface between the single-crystal bulk GaN substrate and the MBE-grown epilayers. The absence of visible features at the nucleation interface, and absence of extended defects in the epilayers indicates successful homoepitaxial growth of GaN *p-n* diodes by MBE.

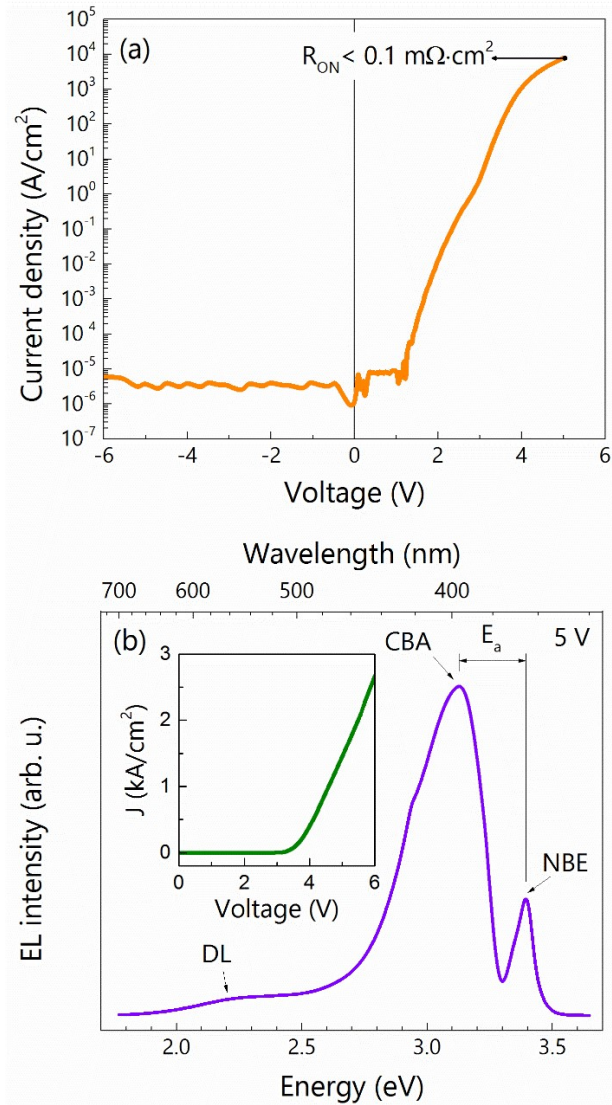

Fig. 2 (a) Current density vs voltage characteristics of the N-polar GaN single-crystal diodes in semilog scale showing high rectification ratio and low on resistance. (b) Electroluminescence spectrum of an N-polar GaN *p-n* diode measured at a forward bias voltage of 5 V, with two major interband and conduction band to acceptor recombination are seen. The inset in (b) shows the current vs forward voltage curve of the N-polar GaN *p-n* diode.

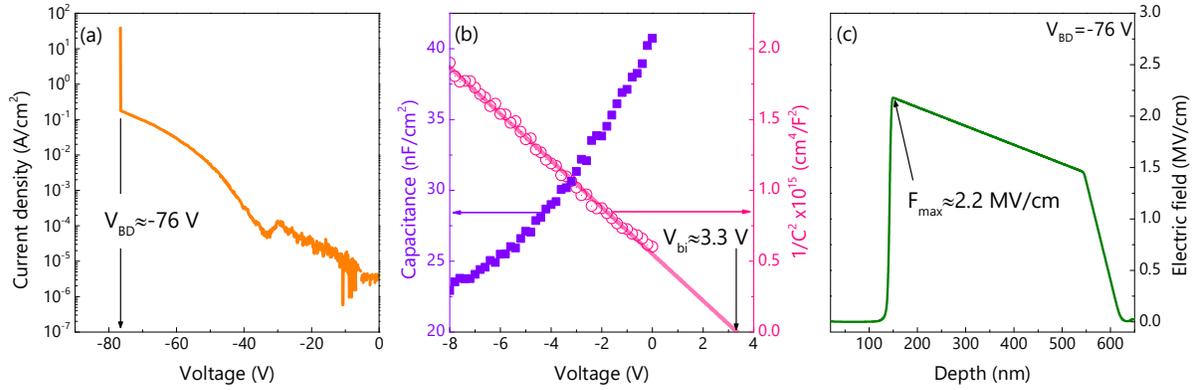

Fig. 3 (a) Semilog plot of reverse-bias current density vs reverse-bias voltage characteristics until breakdown for the N-polar GaN *p-n* diodes. (b) Measured junction capacitance-voltage characteristics from the *p-n* diodes. The *1/C$^2$* vs voltage relation is plotted, and the linear fitting extracts a doping concentration of $9.6\times10^{16}$ cm$^{-3}$ in the unintentionally doped 400-nm-thick GaN of the *p-n* junction. (c) Simulated electric field profile along the vertical direction of the *p-n* junction calculated with the breakdown voltage and doping density, indicating the peak electric field at the *p/n* junction.